\documentstyle[11pt,jd24,epsf]{article}

\begin{document}

\title{Pulsating Variable Stars in the MACHO Bulge database: The Semiregular
Variables}

\author{D. Minniti$^1$, C. Alcock, D. Alves, K. Cook, S. Marshall}
\affil{Lawrence Livermore National Laboratory}
\author{R. Allsman, T. Axelrod, K. Freeman, B. Peterson, A. Rodgers}
\affil{Mount Stromlo and Siding Springs Observatory}
\author{K. Griest, T. Vandehei, A. Becker, C. Stubbs, A. Tomaney}
\affil{University of California San Diego, University of Washington}
\author{D.Bennett, M. Lehner, P.Quinn}
\affil{Notre Dame, Sheffield University, European Southern Observatory}
\author{M. Pratt, W.Sutherland, D.Welch}
\affil{MIT, Oxford University, McMaster University}
\author{{\it (The MACHO Collaboration)}}

\altaffiltext{1}{Invited review for IAU-JD24, to appear in: Pulsating Stars -- Recent Developments in Theory and Observation, 1997, eds. M. Takeuti \& D. Sasselov, (Universal Academy Press: Tokyo).}

\begin{abstract}
We review the pulsating stars contained in the top 24 fields of the
MACHO bulge database,
with special emphasis on the red semiregular stars.
Based on period, amplitude and color cuts, we have selected a sample
of 2000 semiregular variables with $15<P<100$ days.
Their color-magnitude diagram is presented, and 
period-luminosity relation is studied, as well as their spatial 
distribution. We find that they follow the bar, unlike the RR Lyrae
in these fields.
\end{abstract}


\section{Introduction}

Current microlensing experiments (MACHO, Alcock et al. 1997;
EROS, Aubourg et al. 1995; OGLE, Udalski et al. 1993; DUO, Alard 1996)
have produced exquisite light curves of variable stars as
byproducts. These thousands of light curves allow us to address 
advanced questions regarding stellar evolution,
pulsational physics, the distance scale and Galactic structure, 
which would in turn aid in the interpretation of microlensing.
This paper reports on the MACHO pulsating variable stars, with emphasis on
the semiregular variables in the bulge, which have not been previously analyzed.
The LMC counterparts are discussed in this book by Alves at al. (1998), and 
a list of references regarding other variable stars in the MACHO database
can be found in our web page at {\bf http://wwwmacho.mcmaster.ca/}.
This paper is part of a systematic study of the structure of the inner
Milky Way using different distance indicators contained in this
database (Minniti et al. 1996, 1997).

A description of the MACHO system is given by Cook et al. (1995, 1996). Briefly,
the 1.27m telescope at MSSSO obtains nightly images of several bulge,
LMC, or SMC fields in two passbands simultaneously.
These images are photometered on-line, and later calibrated to produce
accurate $V$ and $R$ magnitudes. Variable stars are also identified 
automatically, and their light curves are phased using a 
super-smoother algorithm (Cook et al. 1995).

The period-amplitude
diagram of about 50000 periodic variables in the top 24 MACHO fields from the
1993 season is shown in Figure 1.  
These variables have periods $0.1<P<100$ days.
Stars with periods longer than 100 days (e.g. Miras) have not been phased yet,
although our light curves in the bulge contain typically 700 points over
a 5-year baseline.  The presence of RR Lyrae type ab
at $0.4<P<1$ day and semiregulars at $15<P<100$ days is evident.
In addition, there is a large number of eclipsing binaries
with $A<1$, spanning a wide range of periods. The binaries with shorter periods
overlap in the diagrams of Figure 1 with the RR Lyrae type c.

The nightly observations introduce aliasing when phasing the light curves
of variable stars. These aliased periods with $P=1/n$ days (with $n=1, 2, ...$)
are seen in Figure 1. To eliminate aliases, we demand that the
periods in both bands are the same to within a few percent, i.e.
$P_V = P_R \pm 2$\%.

\section{Selection of Pulsating Variable Stars}

Clearly, the visual inspection of the light curves of
50000 bulge variables would be time consuming.
Fortunately, the automatic classification of variable stars can be
very much improved having observations in two different passbands.
The discrimination between eclipsing and pulsating variables in this 
database is most easily done using the amplitude ratios. Most 
pulsating variable stars have $A_R < 0.8 ~A_V$, while the light curves
of most eclipsing binaries have $A_V = A_R$. 

Figure 2 shows the same diagrams as Figure 1, but containing only
pulsating variables, simply selected using 
$P_V = P_R \pm 2$\% and $A_R < 0.8 ~A_V$.
These diagrams look much cleaner, showing the different families of 
pulsating stars.
These are $\delta$ Scuti stars,
with $P<0.2$ days, RR Lyrae stars type c, with $0.2<P<0.4$ days,
RR Lyrae type ab, with $0.4<P<1.1$ days, and semiregular variables,
with $15<P<100$ days.  

\begin{figure}
\plotone{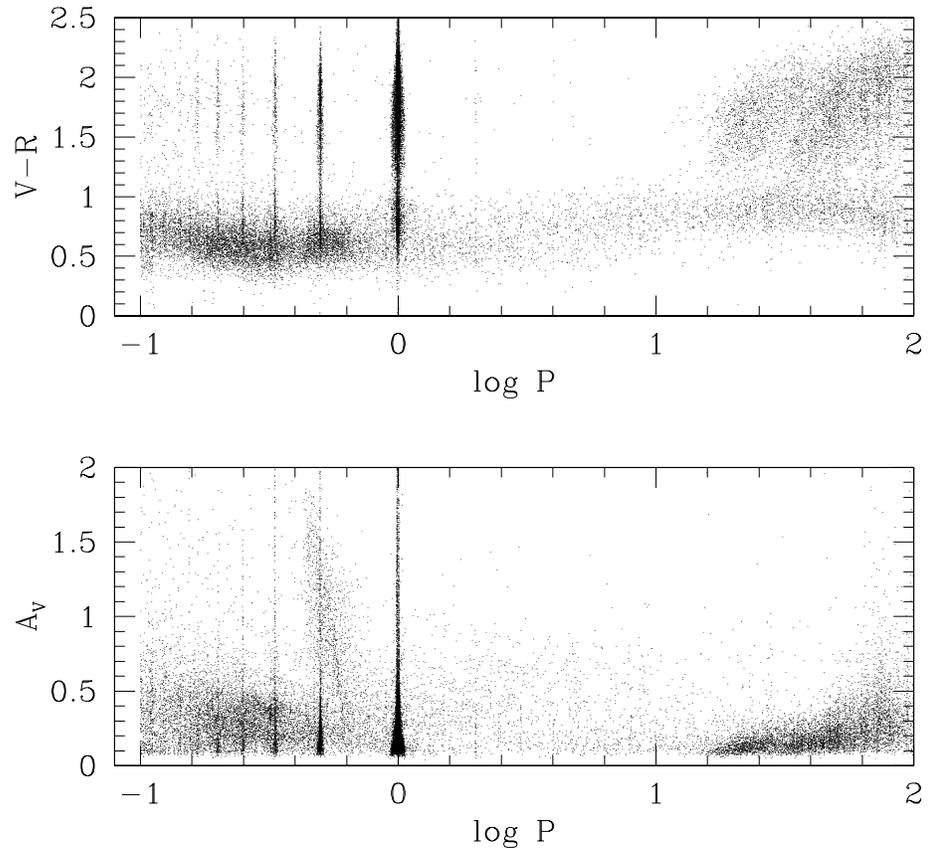}
\caption{
Period-amplitude  and period-color diagrams for periodic bulge stars
with $0.1<P<100$ days.
This includes all pulsating and eclipsing variables in the
top 24 bulge fields observed by MACHO.
The vertical groups of points at submultiples of 1 day are due to
aliasing. 
}
\end{figure}

\begin{figure}
\plotone{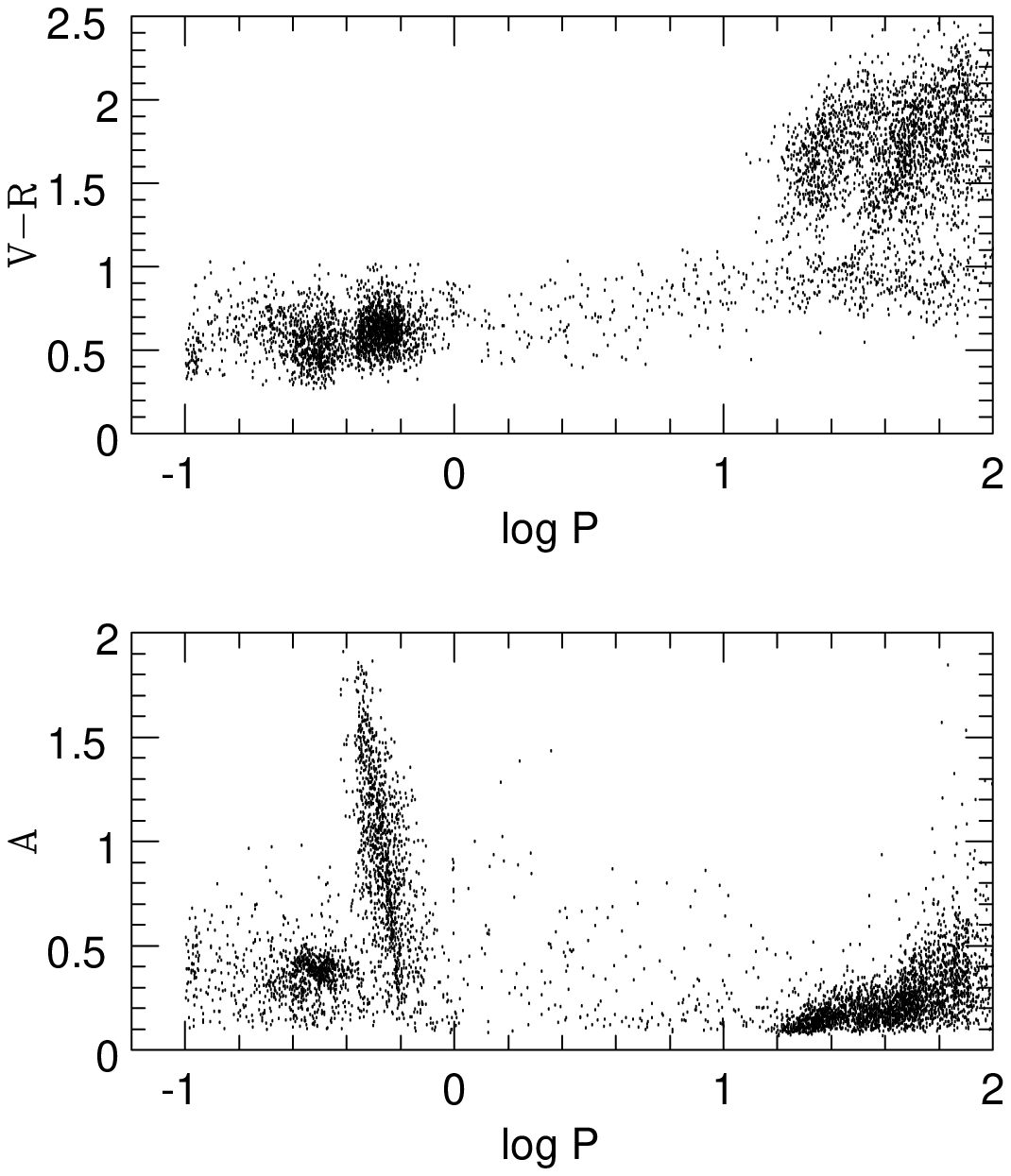}
\caption{Same as Figure 1, showing only
bulge pulsating stars with $0.1<P<100$ days, selected using the cuts
$P_R = P_V \pm 2$\%, and $A_R<0.8 ~A_V$.
There are, in order of increasing $log ~P$: $\delta$ Scutis, RR Lyrae type c, 
RR Lyrae type ab, and semiregular variables. Note the absence of Cepheids,
with $1<P<15$ days, which are very numerous in the LMC.
}
\end{figure}

\subsection{The RR Lyrae Stars}

The bulge RR Lyrae in the MACHO database were studied by Minniti et al. (1996).
They found that (1) they do not follow the bar in the inner regions,
(2) they are very concentrated, (3) the fainter RR Lyrae belong to the
Sgr dwarf galaxy, located behind the bulge, and (4) the period-amplitude 
diagram differs from that of the LMC, with larger fraction of RRc/RRab types.

\subsection{The $\delta$ Scuti Stars}

The large amplitude $\delta$ Scuti stars in the MACHO bulge database
were studied by Minniti et al. (1997). They found that (1) these 
stars belong to the bulge, and not to the disk, (2) they are
most likely bulge blue stragglers, (3) they are potentially good 
distance indicators, and (4) they are also very concentrated.

\subsection{The Cepheid Stars}

Cepheid variable stars are very numerous in the Magellanic Clouds (Alcock et al. 1995,
Welch et al. 1996, Sasselov et al. 1997). Note, however, that 
there are very few Cepheids candidates in
the bulge database (Figure 2). These few Cepheids, with $1<P<15$ days,
and $A<1.5$, belong to the disk of our galaxy. The bulge population is
too metal rich and too old to contain such stars. 

\begin{figure}
\plotone{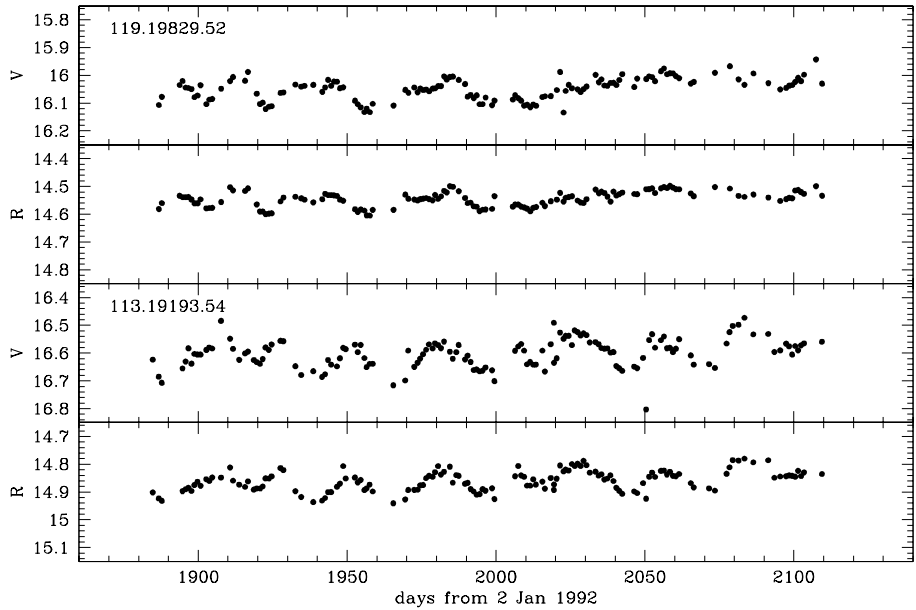}
\caption{Observed $V$ and $R$ light curves for the 1997 season
of two bulge semiregular variables. These correspond to the shorter period
sequence: Macho 119.19829.52, with $P=21.5$ days (top panels), 
and Macho 113.19193.54, with $P=26.0$ days (bottom panels).
}
\end{figure}

\section{The Semiregular Variables}

The rest of this paper will be devoted to bulge semiregular variable stars.
Figures 3 and 4 show the MACHO light curves of 4 such stars, illustrating
why they are called semiregular variables.
Semiregulars are red giant pulsators, more numerous that Miras,
but with smaller amplitudes and shorter periods. They have, however,
been less well studied than the Miras (see Gautschy \& Saio 1996, 
Whitelock 1996, Percy et al. 1996).

As mentioned above, pulsating bulge variables are selected using
cuts in periods and amplitudes: 

\noindent $\bullet$ $P_V=P_R \pm 2$\% and $A_R < 0.8 ~A_V$.

The bulge semiregular variables are chosen using further cuts in
periods, colors and amplitudes:

\noindent $\bullet$ $P>15$ days to discard 
shorter period pulsators ($\delta$Sct, RR Lyr, Cepheids),

\noindent $\bullet$ $P<100$ days to eliminate variables with longer periods (Miras, LPVs),

\noindent $\bullet$ $V-R>1.0$ to discriminate from
bluer variables, 

\noindent $\bullet$ $A_V>0.1$ to ensure good quality light curves,  and

\noindent $\bullet$ $A_V<1.0$ to avoid
the occasional Mira or LPV that has been wrongly phased.

These cuts produce a sample of 2000 semiregular variables in the bulge.

Note that the periods used in this paper were determined from the 1993 data
only. The long term period stability of these stars in not known.
Some of these periods may change from season to season, as a result of the
semiregular behavior of these variables. For example, multiple and variable
periods have been observed in some semiregulars (e.g. Mantegazza 1988,
Zsoldos 1993, Lebzelter et al. 1995).  However, the selection criteria
applied to our data guarantees that during at least
one season, the periods measured using both passbands agree.

\begin{figure}
\plotone{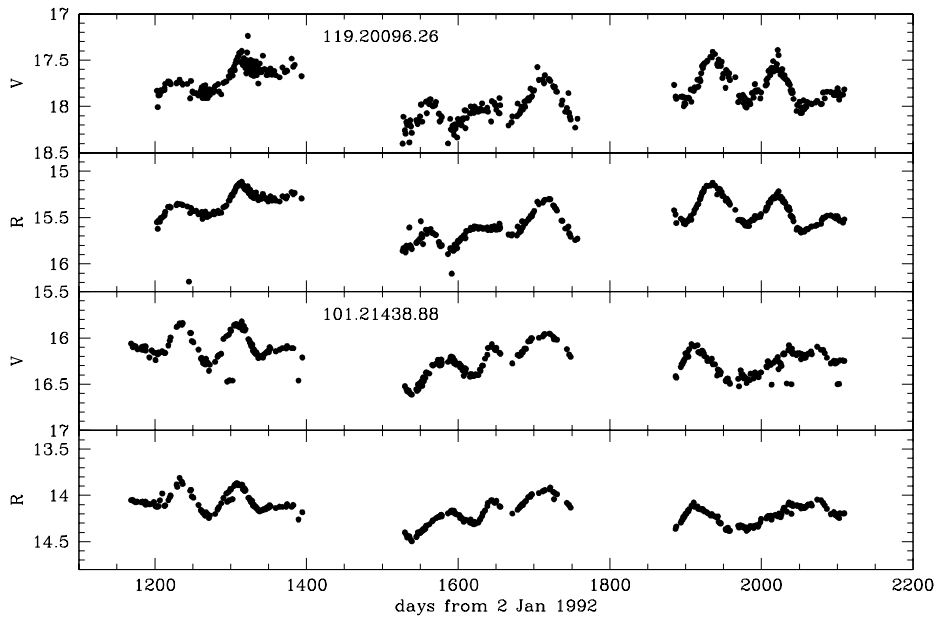}
\caption{Observed $V$ and $R$ light curves for the 1995, 1996 and 1997 seasons 
of two bulge semiregulars. These belong to the longer period
sequence: Macho 119.20096.26, with $P=72.5$ days (top panels), 
and Macho 101.21438.88, with $P=80.6$ days (bottom panels).
}
\end{figure}

The selected sample of 2000 bulge semiregulars is plotted in Figure 5.
The onset of pulsation occurs at $P=15$ days, with low amplitudes, $A=0.1$.
We have not considered variables with $A<0.1$, so there may be
more lower amplitude semiregulars with periods $P<15$ days.
The amplitudes increase with increasing periods, but they are not
larger than $A=1$. 

Figure 5 shows that there  are
two sequences of semiregulars in the bulge.
These sequences are better
seen in the period-color diagram, where a clear separation
is obtained applying the cuts
$V-R >2.5 ~log P - 2.12$, and $V-R < 2.5 ~log P -2.12$.
These sequences have stars with with $15<P<40$ days, and $32<P<100$ days,
and both sequences have mean amplitudes increasing with period.
As red giants evolve upwards on the giant branch
(e.g. Vassiliadis \& Wood 1994), and the semiregulars
with $32<P<100$ days should be more evolved than the ones with $15<P<40$ days.
Indeed, the long period semiregulars tend to be redder and brighter in
the mean than the short period ones.

Similar sequences, although much tighter, are seen in the MACHO LMC database
(e.g. Cook et al. 1996, Alves et al. 1998).  We speculate that
these sequences represent stars pulsating in different modes, as predicted
by Wood \& Cahn (1977).
These sequences provide a more natural classification scheme than the 
traditional SRa and SRb classes.
The relation of these sequences to the longer period Mira variables has not
been well studied yet. Unlike these longer period variables,
semiregulars with $P<100$ days do not necessarily have circumstellar shells
(Kerschbaum \& Hron 1996). 
Previous studies have shown that in the Milky Way disk, the kinematics and
scale-heights of the semiregular variables are similar to these of
the Miras (Jura \& Kleinmann 1992), suggesting that they trace similar populations.
In particular, a possible extension of the Mira period-luminosity 
relations into the shorter periods covered by the semiregular variables 
(e.g. Feast 1996),
could be useful for the determination of the extragalactic distance scale.

\begin{figure}
\plotone{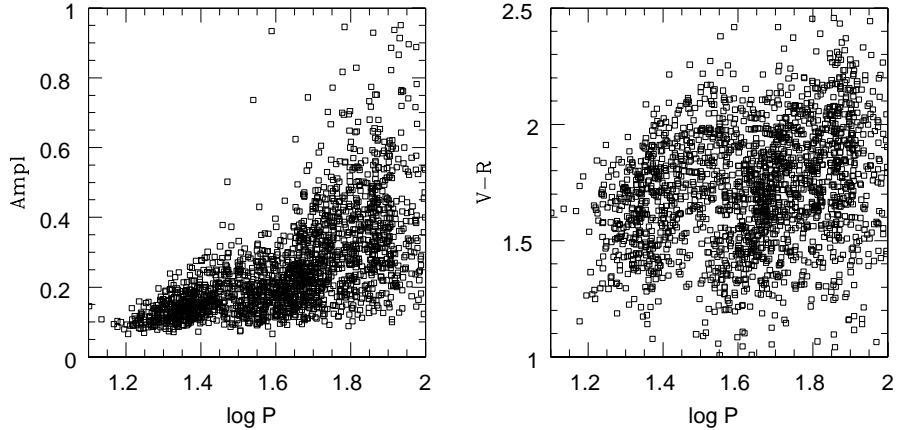}
\caption{
Left panel: Period-amplitude diagram for 2000 bulge semiregular variables.
Right panel: Period-color diagram. Note the
two groups with short and long periods.
}
\end{figure}

\subsection{The Color-Magnitude Diagrams}

Figure 6 shows the location of the 2000 semiregular variables in the
color-magnitude diagram. Overplotted are 10000 stars in Baade's
window, a typical bulge field, showing the disk main sequence at  around
$V-R = 0.5$, the bulge red giant branch at $V-R = 0.7-1.0$, and the 
red giant branch clump at about $V=17$, $WV=14$, $V-R = 0.9$.
The reddening independent magnitude WV is defined as 
$WV= V-3.97 (V-R)$. 

These diagrams illustrate that the
semiregular variables are located in the bulge, and not in the
foreground disk. 
Among the reddest stars in the bulge, they are oxygen rich giants
located at the tip of the bulge red giant branch.
The semiregulars lie along the direction of the reddening vector in the
color-magnitude diagram, mostly due to the
large and non-uniform insterstellar absorption in the
observed bulge fields (note that circumstellar extinction should be reduced in
the semiregulars compared with the longer period Mira variables).
The MACHO bulge images with exposure times of 150 sec
saturate at $V \sim 13$, so a few of the brightest stars
may have been missed.

\begin{figure}
\plotone{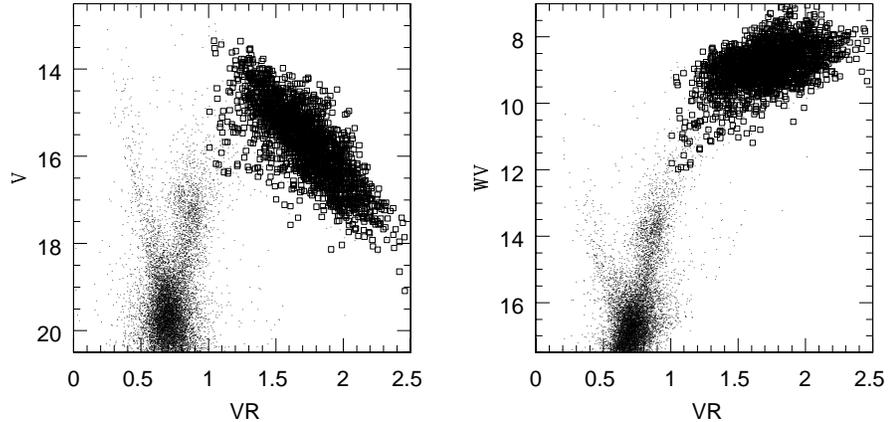}
\caption{Color-magnitude diagrams 
of 2000 bulge semiregular variables (squares),
overplotted with 10000 stars in Baade's window,
a typical bulge field (points). 
Left: $V$ $vs$ $V-R$.
Right: Reddening independent magnitude $WV$ $vs$ $V-R$ diagram. 
}
\end{figure}

\subsection{The Period-Luminosity Relation}

Semiregular variables are bright, easy to detect, and
very numerous in old stellar populations.
They are fainter than Miras, but more numerous. They are brighter and
more numerous than RR Lyrae and $\delta$ Scuti stars. Their luminosity
is roughly comparable with Cepheids, which are bluer, but they are also
much more numerous.  It would then be important if they could
be used as distance indicators. Is there a period-luminosity relation
for these stars? This question can be addressed with the MACHO bulge and
LMC database, that contains thousands of semiregulars, located
at about the same distance.

Figure 7 shows the period-luminosity plane for these variables.
The spread in the $V$ magnitudes 
is due to the bulge line-of-sight depth, and to differential
reddening. Due to this scatter, no clear dependence of $V$ with $P$ is
observed. We then use the reddening independent $WV$ magnitudes,
which show smaller spread, solely due to the bulge
line-of-sight depth, and also a clearer separation between the two sequences
discussed above.

There appears to be another sequence about 
2 magnitudes fainter, containing fewer stars
than the two principal sequences. This third sequence is
seen in the right panel of
Figure 7 for $P>30$ days, and $10<WV<12$. These stars could be 
semiregulars in the Sgr dwarf galaxy, located behind the bulge, at a distance 
of 25 kpc from the Sun (see Alard 1996 and Alcock et al. 1997). 
However, a similar sequence may also be present in the LMC, so the 
identification and interpretation remains unclear.

The right panel of Figure 7 shows the dependence of luminosity on
the period of the stars. The inferred period-luminosity relation for
the whole sample of semiregulars with $15<P<100$ days is:
$$WV \approx -1.78 ~log~P + 11.56$$
However, dividing into two groups, presumably representing variables
pulsating in different modes, yields the following preliminary
period-luminosity relations:
$$WV \approx -3.33 ~log~P + 13.66 ~~~(15<P<40 ~d)$$
$$WV \approx -3.33 ~log~P + 14.47 ~~~(32<P<100 ~d)$$
Followup observations of these variables in the near-infrared would be
very useful, in order to obtain much tighter
period-luminosity relations (see Feast 1996).

\begin{figure}
\plotone{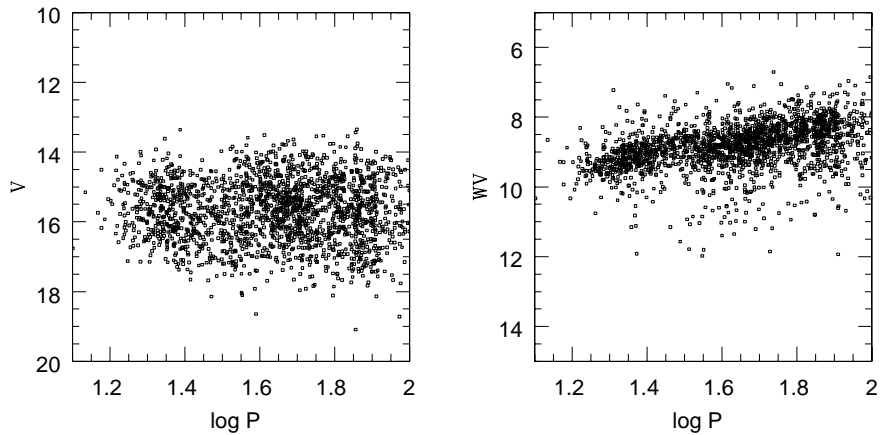}
\caption{
Left panel: Period-$V$ magnitude relation for 2000 semiregular variables.
Right panel: Period-$WV$ magnitude relation.
}
\end{figure}

\subsection{The Spatial Distribution}

We have computed the mean reddening independent magnitudes of the 
semiregulars for each of the top 24 MACHO bulge fields.
These mean magnitudes are plotted as function of Galactic longitude in Figure 8.
This figure shows that the bulge semiregular variables trace the bar:
the semiregulars away from the Galactic minor axis appear to be
brighter, and therefore closer, than the ones along the minor axis.
The dependence of the mean $V$ and $R$ 
magnitudes on Galactic longitude is similar to that of the mean
$WV$ magnitudes shown in Figure 8, with somewhat larger scatter due to variable
reddening.

The trend seen in Figure 8 follows that of the bulge clump giants
(Stanek et al. 1996,
Alcock et al. 1997), and of the Mira variables (Whitelock 1993),
which also clearly show a barred distribution.
This suggests that the semiregular variables are metal-rich like the bulk of
the bulge population,
because metal-poor stars (such as RR Lyrae) do not follow the bar 
(Alcock et al. 1997).

\begin{figure}
\plotone{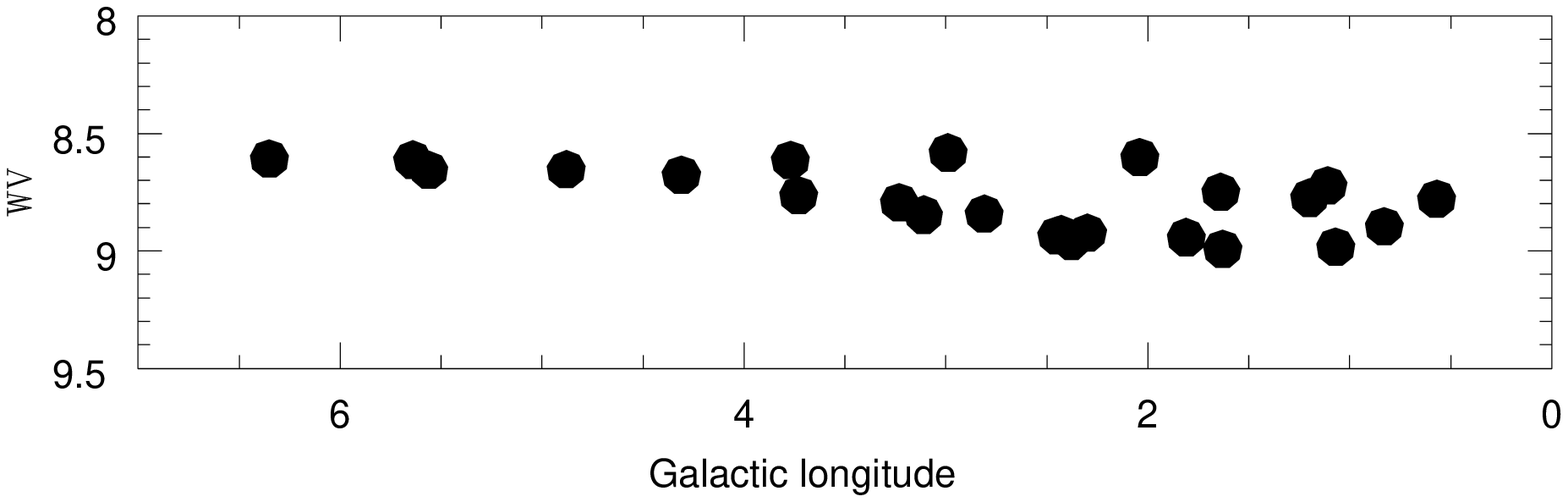}
\caption{Mean reddening independent magnitude for 
bulge semiregular variables in each of the top 24 MACHO fields
$vs$ Galactic longitude.
Each point represents the mean magnitude in one field containing between 30
and 250 objects.  This figure shows that the semiregulars 
follow the barred distribution seen in the clump giants and other tracers.
}
\end{figure}

\subsection{Bulge Semiregular Variables: New Challenges}

Past studies of semiregular variables have not been as detailed as those of other
pulsating stars like $\delta$ Scuti, RR Lyrae, Cepheids or Miras.
This was partly due to the semiregular nature, which makes it necessary to 
have observations over extended periods of time.
Nonetheless, carefully selected samples of these
variables can be used as distance indicators, with the advantage of being
more numerous than some of these other variables.

We have shown that there is a period-luminosity relation for semiregular
variables in the bulge, and that they can be used as tracers of the structure 
of the inner Milky Way. In particular, we find that 
they follow the inner bar, like clump giants, and Mira variable stars.

There are, however, several questions that remain open for further 
study concerning the bulge semiregular variables:

-- How do they relate to the Mira-type variables with $P>100$ days?

-- What are their pulsation modes?

-- What is their long term pulsational stability? 

-- Is there a metallicity dependence of their period-luminosity relation?

The current and future microlensing experiments would help solve these
issues in the near future. A complete inventory of semiregular variables
would also aid in the selection of real microlensing events in more
distant galaxies using difference image photometry (e.g. 
Columbia-VATT, Tomaney \& Crotts 1997;
AGAPE, Ansari et al. 1997).

\acknowledgments
We are grateful for the skilled support by the technical staff at MSSSO.
Work at LLNL is supported by DOE contract W7405-ENG-48.
Work at the CfPA is supported NSF AST-8809616 and AST-9120005.
Work at MSSSO is supported by the Australian Department of Industry,
Technology and Regional Development.
WJS is supported by a PPARC Advanced Fellowship.
KG thanks support from DOE OJI, Sloan, and Cottrell awards.
CWS thanks support from the Sloan, Packard and Seaver Foundations.


\begin{references}
\reference Alard, C. 1996, ApJ, 458, L17
\reference Alcock, C. et al. (The MACHO Collaboration), 1995, AJ, 109, 1653
\reference Alcock, C. et al. (The MACHO Collaboration), 1997, ApJ, 474, 217
\reference Alcock, C. et al. (The MACHO Collaboration), 1998, ApJ, in press
\reference Ansari, R., et al. (The AGAPE Collaboration), 1997, A\&A, 324, 843
\reference Alves, D., et al. (The MACHO Collaboration), 1998, this book
\reference Aubourg, E., et al. (The EROS Collaboration), 1995, A\&A, 301, 1
\reference Beaulieu, J.-P., et al. (The EROS Collaboration), 1997, A\&A, 321, L5
\reference Cook, K., et al. (The MACHO Collaboration), 1995, in ASP No. 83, p. 221
\reference Cook, K., et al. (The MACHO Collaboration), 1996, in ``Variable Stars and the Astrophysical Returns of Microlensing", eds. R. Ferlet et al. (Editions Frontieres: Paris), p. 17
\reference Feast, M. W. 1996, MNRAS, 278, 11
\reference Gautschy, A., \& Saio, H. 1996, ARA\&A, 34, 551
\reference Jura, M., \& Kleinmann, S. G. 1992, ApJS, 79, 105
\reference Kerschbaum, F., \& Hron, J. 1996, A\&A, 308, 489
\reference Lebzelter, T., Kerschbaum, F., \& Hron, J. 1995, A\&A, 298, 159
\reference Mantegazza, L. 1988, A\&A, 196, 109
\reference Minniti, D., et al. (The MACHO Collaboration), 1996, in ``Variable Stars and the Astrophysical Returns of Microlensing", eds. R. Ferlet et al. (Editions Frontieres: Paris), p. 257
\reference Minniti, D., et al. (The MACHO Collaboration), 1997, in IAU Coll. 189 on ``Fundamental Stellar Properties: The Interaction Between Observation and Theory", eds. T.R. Bedding et al. (Kluwer: Dordrecht), p. 293
\reference Percy, J. R., et al. 1996, PASP, 108, 139
\reference Sasselov, D., Beaulieu, J.-P., et al. (The EROS Collaboration), 1997, A\&A, 324, 471
\reference Stanek, K. Z., et al. (The OGLE Collabotarion), 1994, ApJ, 429, L73
\reference Tomaney, A. B., \& Crotts, A. P. S. 1997, AJ, 112, 2872
\reference Udalski, A., et al. (The OGLE Collabotarion), 1993, Acta Astr., 43, 69
\reference Vassiliadis, E., \& Wood, P. R. 1994, ApJS, 92, 125
\reference Welch, D., et al. (The MACHO Collaboration), 1996, in ``Variable Stars and the Astrophysical Returns of Microlensing", eds. R. Ferlet et al. (Editions Frontieres: Paris), p. 205
\reference Whitelock, P., 1993, in IAU Symp. 153 on "Bulges of Galaxies",
eds. H. Dejonghe \& H. Habing (Kluwer: Dordrecht), p. 39
\reference Whitelock, P. 1996, in ``Variable Stars and the Astrophysical Returns of Microlensing", eds. R. Ferlet et al. (Editions Frontieres: Paris), p. 163
\reference Wood, P. R., \& Cahn, 1977, ApJ, 211, 499
\reference Zsoldos, E. 1993, A\&A, 268, 149
\end{references}
\end{document}